\documentclass[aps,prb,twocolumn,superscriptaddress]{revtex4-1}
\usepackage{graphicx}
\usepackage{dcolumn}
\usepackage{bm}
\usepackage{amsmath}
\usepackage{amssymb}
\usepackage{color}

\begin{document}

\title{Nonequilibrium Membrane Dynamics Induced by  Active Protein Interactions and Chemical Reactions: A Review}

\author{Hiroshi Noguchi}
\email[]{noguchi@issp.u-tokyo.ac.jp}
\affiliation{Institute for Solid State Physics, University of Tokyo, Kashiwa, Chiba 277-8581, Japan}

\begin{abstract}
Biomembranes wrapping cells and organelles
are not only the partitions that separate the insides
but also dynamic fields for biological functions
accompanied by membrane shape changes.
In this review, we discuss the spatiotemporal patterns and fluctuations of membranes under nonequilibrium conditions. 
In particular, we focus on theoretical analyses and simulations.
Protein active forces enhance or suppress the membrane fluctuations;
the membrane height spectra are deviated from the thermal spectra.
Protein binding or unbinding to the membrane is activated or inhibited 
by other proteins and chemical reactions, such as ATP hydrolysis.
Such active binding processes can induce traveling waves, Turing patterns, and
membrane morphological changes.
They can be represented by the continuum reaction-diffusion equations 
and discrete lattice/particle models with state flips.
The effects of structural changes in amphiphilic molecules on the molecular-assembly structures
are also discussed.
\end{abstract}

\maketitle

\section{Introduction}
\label{introduction}

Cells and most organelles are enclosed in membranes consisting of lipids and proteins.
The membrane separates the solutions inside and outside.
Inner ion concentrations are controlled by ion channels on the membrane.
Membranes dynamically change their shapes through interacting with membrane proteins and the cytoskeleton.
Lipids and proteins are transported by small vesicles that divide from plasma or organelle membranes and fuse other membranes.
Chemical waves on membranes are widely observed.\cite{beta17,bail22,gov18,wu21,alla13} 
Understanding the membrane dynamics in nonequilibrium systems is important.

The static properties of lipid membranes have been extensively studied and are well-understood.
The morphology of giant unilamellar vesicles (GUVs) is well-understood based on the bending energy 
with constraints.\cite{lipo95,seif97,svet14,saka12}
Phase separation of ordered and disordered lipid domains has been well investigated
in three-component membranes.\cite{honn09,baum03,veat03,yana08,chri09}
Several protein types are known to regulate membrane curvatures.\cite{mcma05,masu10,baum11,mim12a,suet14,joha15}
These proteins bend the bound membrane and also sense the membrane curvature (i.e., concentrated on the membranes of their preferred curvatures).\cite{baum11,has21,prev15,tsai21} Their curvature dependences are reproduced by mean-field theories.\cite{nogu22a,nogu23b}

Numerical simulations are useful tools for investigating detailed dynamics and examining the essential factors that control dynamics.
Membrane dynamics are ranged from the $n$m to $\mu$m scales; hence, 
various types of membrane models  have been developed from atomistic and coarse-grained molecules to curved surfaces.
Since these models are reviewed in Refs.~\citenum{muel06,vent06,nogu09,marr09,shin12,peze21,dasa24},
 we do not describe model details here. We focus on the dynamics obtained by these simulations.

In this review, we discuss recent studies on nonequilibrium membrane dynamics,
particularly theoretical and simulation studies.
We consider the nonequilibrium conditions caused by protein activity and chemical potential difference.
Hence, we do not discuss nonequilibrium membrane dynamics in external flows\cite{fedo13,lano16,beri21} here.
In Sec.~\ref{sec:flu}, we describe non-thermal membrane fluctuations.
Theoretically, we mainly treat the membrane as a homogeneous surface.
In Sec.~\ref{sec:bind}, we describe the binding and unbinding of curvature-inducing proteins and other molecules to membranes,
which can deviate from the equilibrium balance by the chemical potential difference between 
the solutions on the two sides of the membrane and via activation through chemical reactions.
We consider binding/unbinding of a single protein type.
We also discuss the formation of membrane structures by the fusion of small vesicles.
In Sec.~\ref{sec:reac}, we describe the binding and unbinding of multiple molecules,
which are activated or inhibited by other molecules.
The time evolutions of the molecular concentrations on the membrane are represented by reaction-diffusion equations.
We discuss the coupling of membrane deformation and reaction waves.
In Sec.~\ref{sec:mol}, we describe structure formations induced by chemical reactions that change molecule amphiphilicity.
We discuss structural changes, such as a micelle-bilayer transition, and morphological changes in vesicles.
A summary and outlook are presented in Sec.~\ref{sec:sum}.

\section{Membrane Fluctuations}\label{sec:flu}

In thermal equilibrium,
the bending energy of a membrane in a fluid phase 
is expressed as\cite{canh70,helf73}
\begin{equation} \label{eq:hel}
F_{\mathrm{cv}}= \int \Big[\frac{\kappa}{2} (2H-C_0)^2 + \bar{\kappa} K \Big]\ {\mathrm{d}}A,
\end{equation} 
where the integral is taken over the membrane surface $A$.
$H=(C_1+C_2)/2$ and $K=C_1C_2$ are the mean and Gaussian curvatures,
where $C_1$ and $C_2$ are the principal membrane curvatures.
$C_0$ is the spontaneous curvature and symmetric membranes have $C_0=0$.
Lipid membranes have a bending rigidity $\kappa$ in the range of $\kappa = 10$--$100k_{\mathrm{B}}T$\cite{kara23,dimo14,mars06,rawi00}
and a saddle-splay modulus $\bar{\kappa} \simeq -\kappa$ (also called the Gaussian modulus).\cite{hu12}
The integral of the Gaussian curvature is constant for a fixed membrane topology (Gauss--Bonnet theorem).
The fluctuations of a planar membrane 
are expressed as\cite{safr94,helf84}
\begin{equation} \label{eq:hqeq}
\langle |h(q)|^2 \rangle= \frac{k_{\mathrm{B}}T}{\kappa q^4 + \gamma q^2},
\end{equation}
where  $h(q)$ is the Fourier transform of membrane height,
$k_{\mathrm{B}}T$ is the thermal energy, and
 $\gamma$ is the mechanical surface tension\cite{shib16}
conjugated to the projected membrane area.
For vesicles\cite{helf86,miln87} and membrane tubes,\cite{four07,shib11} the spectra are slightly modified for spherical and cylindrical coordinates, respectively.
The bending rigidity can be determined from the fluctuations of planar membranes in simulations\cite{goet99,lind00,marr01,boek05,bran11,shib11}
and from the fluctuations of GUVs in experiments.\cite{kara23,dimo14,mars06}

In nonequilibrium,
membrane fluctuations can differ from the equilibrium form expressed by Eq.~(\ref{eq:hqeq}).\cite{turl19,pros96,rama00,lenz03,gov04,lawr06}
Theoretically, different exponents such as $\langle |h(q)|^2 \rangle \sim q^{-5}$ and $q^{-1}$ were derived depending on specific conditions.\cite{pros96,gov04}
Experimentally, fluctuations have been observed to have enhanced for red blood cells (RBCs)\cite{turl19,park10,rodr15,turl16,bisw17}
and GUVs containing protein pumps (rhodopsin,\cite{mann01} Ca$^{2+}$-ATPase,\cite{gira05} and F$_1$F$_0$-ATPase\cite{alme17}).
Flux through the protein pumps can provide a vertical force on the membrane.
In addition, RBCs have a hexagonal spectrin network attached to the plasma membrane,
while ATP hydrolysis modulates the network binding to the membrane.
Lateral force generated by this network binding/unbinding and 
vertical force generated by protein pumps are considered to be sources of active noises in RBCs.
The ATP depletion reduces the membrane fluctuations.

\begin{figure}
\includegraphics[width=8.6cm]{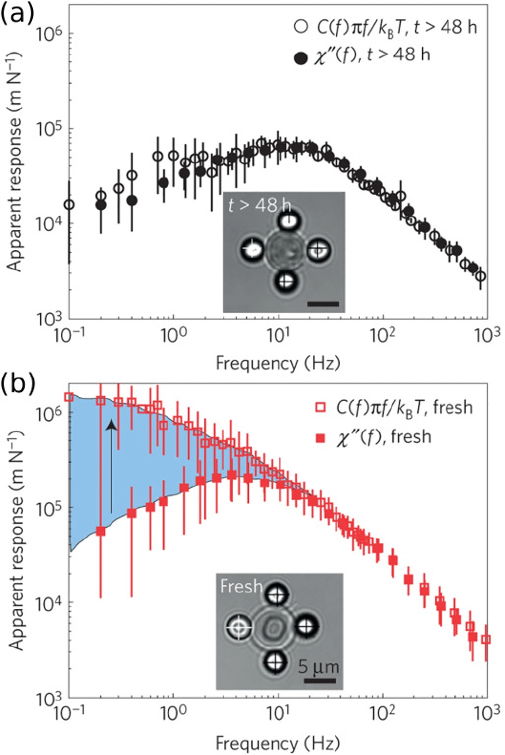}
\caption{
Membrane fluctuations and apparent response 
for (a) RBCs in a glucose-free medium (after 48h) and (b) fresh RBCs.
The filled symbols represent the dissipative response to the applied force.
The open symbols represent the expected equilibrium response calculated from the power spectrum of passive fluctuations by Eq.~(\ref{eq:fdt}).
The insets show the microscope images of RBCs.
Error bars denote standard deviation. Scale bars are $5\,\mu$m.
Reproduced from Ref.~\citenum{turl16} with permission. Copyright 2016, Springer Nature.
}
\label{fig:rbc}
\end{figure}

Turlier et al. reported a violation of the fluctuation--dissipation relation in RBCs.\cite{turl16}
They attached four optically trapped beads to an RBC, as shown in the inset of Fig.~\ref{fig:rbc},
and measured the displacement of one bead that was free under a weak laser trap or moved by a sinusoidal oscillation of the trapping position.
In the former measurement, the power spectrum $C(f)$ of the passive fluctuations is obtained,
where $f$ is the frequency.
In the latter, the complex response function $\chi'(f) + \mathrm{i}\chi''(f)$ is obtained
 as the ratio of the Fourier transforms of the position and force.
In thermal equilibrium, the relation
\begin{equation} \label{eq:fdt}
C(f)= \frac{k_{\mathrm{B}}T}{\pi f} \chi''(f)
\end{equation}
is derived as the fluctuation-dissipation theorem.\cite{call51,kubo66}
Previously, the deviation from this relation has been measured in reconstituted actin-myosin networks.\cite{gnes18,mizu07}
Turlier et al. measured both quantities for ATP-depleted and fresh RBCs, as shown in Fig.~\ref{fig:rbc}.
After a long incubation ($48$h) in a glucose-free medium,
ATP is depleted in RBCs.
Subsequently, the fluctuation-dissipation relation, Eq.~(\ref{eq:fdt}), is satisfied [Fig.~\ref{fig:rbc}(a)].
In contrast, the fresh RBCs exhibit a deviation at low frequencies [blue region in Fig.~\ref{fig:rbc}(b)].
This clearly shows non-thermal fluctuations and rules out the possibility 
of the enhancement being caused by changes in the mechanical properties and structures in equilibrium.
Moreover, these spectra were reproduced by theory and simulations.\cite{turl16}
 
\begin{figure}
\includegraphics[width=8.6cm]{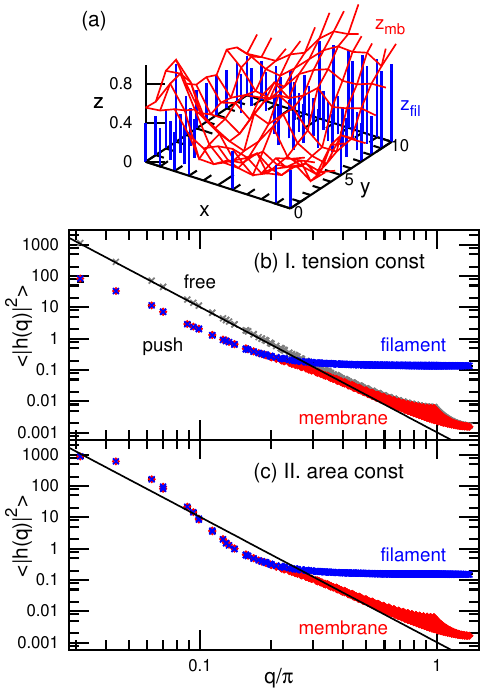}
\caption{
Membrane fluctuations pushed by filament growth.
(a) A snapshot. The membrane is modeled as a red squared mesh,
and filaments are modeled as blue straight bars.
The filament tips stochastically move in the positive $z$ direction,
and the membrane height $z_{\mathrm{mb}}$ is varied by the MC method.
(b)--(c) The spectra of membrane height (red) and filament height (blue).
(b) The surface tension is fixed to be zero. (c) The membrane area is fixed.
The spectrum of the free membrane is also shown in (b) as gray crosses,
which follows the relation $\langle|h(q)|^2\rangle = k_{\mathrm{B}}T/\kappa q^4$ (solid black line).
Reproduced from Ref.~\citenum{nogu21}. Licensed under CC BY.
}
\label{fig:push}
\end{figure}

We have explained the fluctuation enhancement.
Conversely, membrane fluctuations can be suppressed in a moving membrane pushed by protein filament growth.\cite{nogu21}
In crawling cells such as fish keratocytes,
actin filaments grow at the front side of the cells
and push the membrane forward.\cite{svit18,skru20,mogi20,zhao18,okim22}
The filaments grow following a Brownian ratchet mechanism.\cite{dogt05,hawa01,pesk93,mogi99}
When the membrane is separated from a filament via thermal fluctuations,
a protein monomer attaches to the filament.
When the filaments form a network structure, the front membrane has a macroscopically flat shape in lamellipodia.\cite{svit18,skru20}
Fluctuations of the membrane pushed by the filaments are calculated using Monte Carlo (MC) simulations.\cite{nogu21}
The membrane and filament are modeled as a square mesh and vertical rod, respectively [see Fig.~\ref{fig:push}(a)].
For tensionless membranes ($\gamma=0$), the membrane fluctuations are suppressed at long wavelengths (low $q$), as shown in Fig.~\ref{fig:push}(b).
At short wavelengths (high $q$), 
the membrane spectra are not modified from the equilibrium spectrum;
 the filament surface exhibits a flat spectrum of white noise, 
as in the absence of membrane--filament interactions.
In contrast, when the membrane area is constrained, a low $q$ mode is enhanced to compensate for the reduction in the excess membrane area 
due to the fluctuation suppression [see Fig.~\ref{fig:push}(c)].
As the constraint membrane tension increases, membrane fluctuations are reduced in thermal equilibrium [see Eq.~(\ref{eq:hqeq})]; 
thus, the filament growth begins to enhance the fluctuations.
Therefore, the bending energy and tension cause the opposite response to the filament growth.

Fluctuations of nuclear envelopes are also enhanced by the interactions with actin filaments and microtubules
outside of the nucleus.\cite{almo19,bied20}
The contraction by these filaments can induce blebbing of the envelopes.\cite{bied20}
Similar blebbing was also observed in the eryptosis of RBCs\cite{qadr17}
and \textit{in vitro} experiments of liposomes with actomyosin network.\cite{lois16}
The blebbing was theoretically explained as the buckling of locally compressed membrane\cite{sens07} 
and was simulated by a triangular network model of fluid and elastic layers.\cite{span11}

When lipid vesicles involve micro-swimmers inside,
the vesicles exhibit tubules and polyhedral-shape formations.\cite{taka20,vutu20,iyer22,pete21}
The filaments are assembled in the tubules and apexes via direct and membrane-mediated interactions.
The neighboring rod-shaped filaments are aligned by their excluded-volume interactions,
leading to the formation of the membrane apex.\cite{pete21,abau19,lee23}

In living cells, the fusion and fission of vesicles to membranes occur with the help of proteins.
Spherical buds are formed on the membranes after fusion and before fission.
Krishnan and Sunil Kumar modeled them as an addition and removal of a vertex in triangulated membrane
and simulated membrane dynamics.\cite{kris22}
Since a newly added vertex is located at a deviated position from the membrane plane,
the vertex addition produces a high bending energy locally such that the frequent addition destabilizes 
the original spherical shape.

\begin{figure*}
\includegraphics[width=17.4cm]{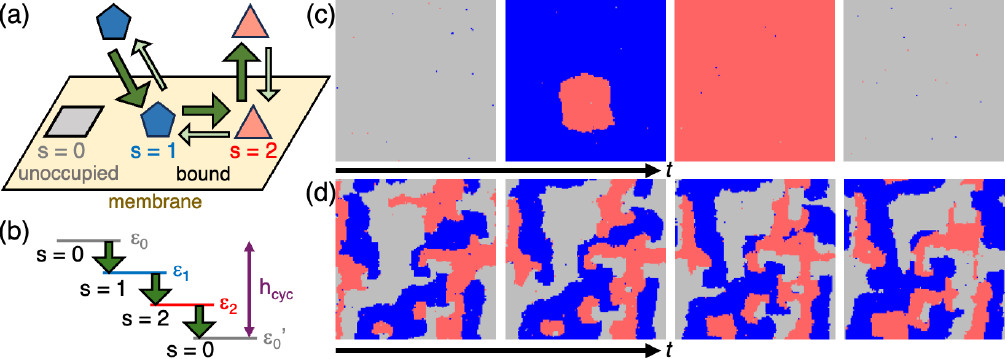}
\caption{
Spatiotemporal patterns in cyclic Potts model.\cite{nogu24a} (a) Schematic of three states on a membane:
an unoccupied state ($s=0$) and two types of bound states ($s=1$ and $2$).
(b) Energy levels. One cycle consumes energy $h_{\mathrm{cyc}}$.
(c)--(d) Sequential snapshots in the cyclically symmetric condition ($\varepsilon_0-\varepsilon_1=\varepsilon_1-\varepsilon_2=\varepsilon_2-\varepsilon_{0}'$).
(c) Homogeneous cycling mode at $h_{\mathrm{cyc}}=2.4$. 
The dominant phases cyclically change as $s=0\to 1 \to 2\to 0$, via nucleation and growth.
(d) Spiral waves at $h_{\mathrm{cyc}}=3.6$.
}
\label{fig:sq}
\end{figure*}

\section{Active Binding of Molecules to Membranes}\label{sec:bind}

In living cells, many types of proteins bind to and unbind from membranes.
In particular, curvature-inducing proteins, such as clathrin and Bin/Amphiphysin/Rvs (BAR) superfamily proteins,
regulate the membrane morphology and generate spherical buds and cylindrical tubules.\cite{mcma05,masu10,baum11,mim12a,suet14,joha15}
These bindings are often activated or inhibited by ATP and GTP hydrolysis.
First, we consider one type of protein that bind to the membrane.
The area fraction $\phi$ of the bound proteins develops as\cite{gout21,nogu22a}
\begin{equation}\label{eq:bind}
\frac{\partial \phi}{\partial t} = \eta_{\mathrm{b}}(1-\phi) -  \eta_{\mathrm{u}}\phi + D\nabla^2 \phi,
\end{equation}
where $\eta_{\mathrm{b}}$ and $\eta_{\mathrm{u}}$ are the binding and unbinding rates, respectively,
$D$ is the diffusion constant, and
$\nabla^2$ is the two-dimensional Laplace-Beltrami operator. 
Homogeneous steady states have $\phi = 1/(1+ \eta_{\mathrm{u}}/\eta_{\mathrm{b}})$.
In thermal equilibrium, this density is determined by the chemical potential $\mu$
as $\eta_{\mathrm{u}}/\eta_{\mathrm{b}} = \exp(-\mu/k_{\mathrm{B}}T)$.
The binding ratios of proteins often depend on membrane curvature.
When a bound protein
has a lateral isotropic bending energy $U_{\mathrm {p}}(H,K)$ and inter-protein interactions are negligible,
the chemical potential becomes $\mu = \mu_0 - U_{\mathrm{p}}(H,K)$,
where $\mu_0$ is the constant term (binding chemical potential).
More proteins bind to the membrane with the preferred curvature (curvature sensing)
at $\partial U_{\mathrm{p}}/\partial H = 0$ and $\partial U_{\mathrm{p}}/\partial H +2H \partial U_{\mathrm{p}}/\partial K = 0$
for cylindrical tubes ($K=0$) and spherical vesicles ($K=H^2$), respectively.\cite{nogu22a,nogu21a}
Binding of crescent proteins, such as the BAR superfamily proteins,
also depends on the protein orientation.
Binding theories for isotropic proteins, crescent proteins, and asymmetric proteins
are described in Refs.~\citenum{nogu21a,nogu21b}, Refs.~\citenum{nogu22,nogu23b}, and Ref.~\citenum{nogu24},
respectively.

In nonequilibrium conditions, protein binding and unbinding rates can deviate from the aforementioned detailed balance.
When the binding (or unbinding) energy of the active process is much larger than the bending-energy difference,
it can be modeled as a uniform rate process independent of membrane curvature.\cite{gout21}
Since the protein states can be altered via energy activation, such as ATP binding and hydrolysis,
the binding and unbinding can obey different chemical potentials.
The condition to form steady domain structures, such as hexagonal and stripe shapes, can be shifted 
by adding such an active process.\cite{gout21,nogu23}
Moreover, the spatiotemporal patterns can also be formed.

As a simple situation, we first describe patterns in which membrane deformation is negligible.
Bound proteins have two states: one that binds strongly on the membrane
and transforms into the other that unbinds more frequently, as shown in Figs.~\ref{fig:sq}(a) and (b).
The same type of states have an attraction with each other to induce a phase separation.
These binding/unbinding dynamics are modeled by a three-state cyclic Potts model.\cite{nogu24a}
When three states are cyclically symmetric ($\varepsilon_0-\varepsilon_1=\varepsilon_1-\varepsilon_2=\varepsilon_2-\varepsilon_{0}'$), homogeneous cycling (HC) and spiral wave (SW) modes appear at low and high activation energy, respectively.
In the HC mode, one of the states dominantly covers the membrane for most of the period;
however, the dominant states cyclically change as $s=0\to 1 \to 2\to 0$ [see Fig.~\ref{fig:sq}(c)].
This phase change stochastically occurs via nucleation and growth.
In the SW mode, spiral-shaped domains spread in a cyclic manner [see Fig.~\ref{fig:sq}(d)].
Small systems exhibit a continuous transition from the HC to SW modes with increasing activation energy via the temporal coexistence of the two modes, whereas large systems exhibit a discontinuous transition. 
Biphasic domains can ballistically move in asymmetric conditions like amoeba locomotion.\cite{nogu24b}
As the activation energy ($\varepsilon_0-\varepsilon_1$) from $s=0$ to $s=1$ increases while keeping the others ($\varepsilon_1-\varepsilon_2=\varepsilon_2-\varepsilon_{0'}$), the period of the $s=2$ dominant phase increases in the HC mode instead of the $s=1$ phase.\cite{nogu24b} This is due to the suppression of the $s=0$ domain nucleation in the $s=2$ phase
by the sequential flips of $s=0 \to 1 \to 2$.
Wave patterns can be observed in other systems, such as predator--prey systems,\cite{szol14,tain94,szab02,reic07} chemical reactions on a catalytic surface,\cite{ertl08,mikh09,goro94,barr20} and
water transport through a liquid-crystalline monolayer.\cite{tabe03}

\begin{figure}
\includegraphics[width=8.6cm]{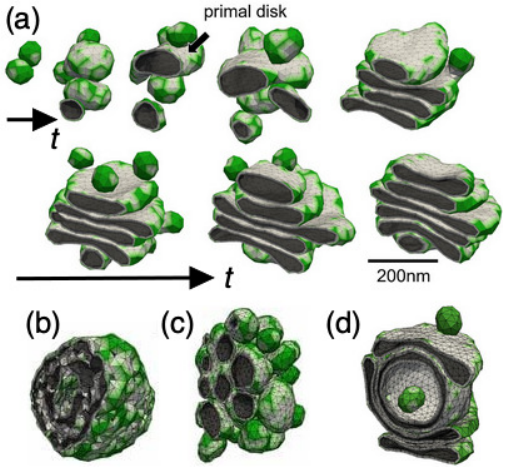}
\caption{
MC simulation of Golgi-apparatus formation.
Green vertices represent membranes bound by curvature-inducing proteins.
(a) Sequential snapshots of the formation process of Golgi-like stacked biconcave vesicles.
Small spherical vesicles fuse into disk-shaped vesicles.
(b)--(d) Various morphologies are obtained depending on conditions.
(a) Reproduced from Ref.~\citenum{tach23}. Licensed under CC BY-NC.
(b)--(d) Reproduced from Ref.~\citenum{tach17} with permission. Copyright (2017) National Academy of Sciences.
}
\label{fig:golgi}
\end{figure}

\begin{figure*}
\includegraphics[width=17.4cm]{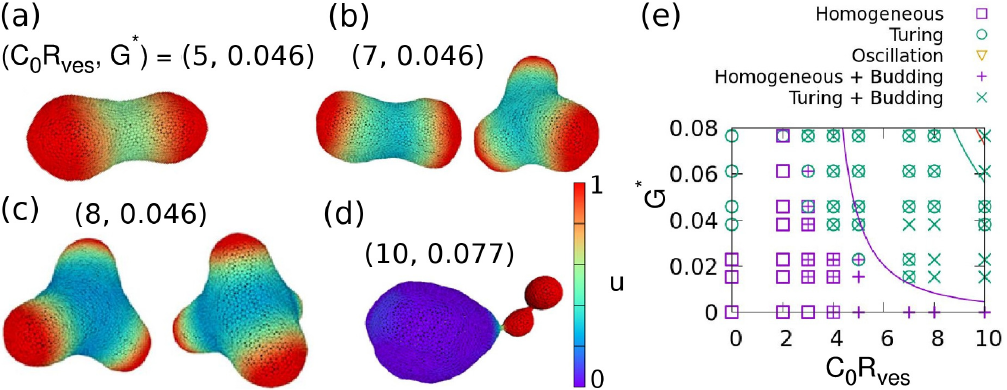}
\caption{
Pattern formation of a vesicle under coupling with reaction-diffusion dynamics
at the reduced volume $V^*=0.8$.
(a)--(d) Snapshots of vesicles at (a) $(C_0R_{\mathrm {ves}}, G^*)=(5,0.046)$, (b) $(7,0.046)$, (c) $(8,0.046)$, and (d) $(10,0.077)$,
where $C_0$ is the spontaneous curvature of proteins and  the normalized coupling constant $G^*= G\kappa/{R_{\mathrm{ves}}}^2$.
(b)--(c) Two different shapes are formed depending on initial states (prolate and discocyte for the left and right snapshots, respectively).
The color indicates the concentration $u$ of the curvature-inducing proteins (see the color bar).
(e) Phase diagram, where
the overlapped symbols indicate the coexistence of multiple patterns.
The purple and green curves represent the Turing and Hopf bifurcations, respectively,
which are analytically determined for a non-deformable spherical vesicle with a radius of $R_{\mathrm {ves}}$.
Reproduced from Ref.~\citenum{tame20}. Licensed under CC BY.
}
\label{fig:turing}
\end{figure*}

In membrane systems, these binding kinetics can be coupled with the membrane deformation.
The curvature can limit domain size, since large domains can form buds or tubules.
Circular domain size can be determined by the balance between the bending energy and surface tension.\cite{gout21}
For vesicles, it is also restricted by vesicle volume.
Tachikawa and Mochizuki simulated the formation of Golgi-apparatus-shaped vesicles using a dynamically triangulated membrane method, as shown in Fig.~\ref{fig:golgi}.\cite{tach17}
They considered the fusion of small vesicles, the binding of curvature-inducing proteins, the membrane--membrane adhesion, and the osmotic pressure between the inside and outside of the vesicles.
By tuning the parameters, such as the osmotic pressure and adhesion energy,
the Golgi-like stacked discocytes are obtained [Fig.~\ref{fig:golgi}(a)].
Under varying conditions, the vesicles can be misfolded into various structures, such as porous multilamellar vesicles [Fig.~\ref{fig:golgi}(b)], 
vesicle aggregates [Fig.~\ref{fig:golgi}(c)], and a stack of stomatocyte and discocytes [Fig.~\ref{fig:golgi}(d)].
Similar misfolded structures have been observed in cell-free reassembly experiments.\cite{rabo95}

\begin{figure}
\includegraphics[width=8.6cm]{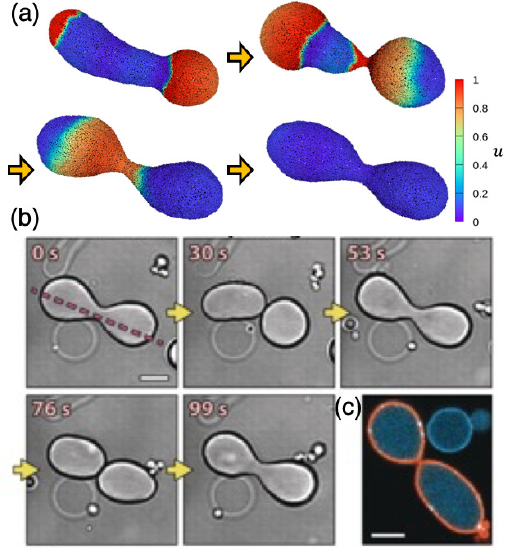}
\caption{
Shape oscillation of a vesicle induced by reaction-diffusion waves.
(a) Sequential snapshots of dynamically-triangulated MD simulation at $V^*=0.65$.
The concentration $u$ of curvature-inducing proteins is indicated by different colors (see the color bar).
(b)--(c) Experiment on GUVs with a Min protein system.
(b) Sequential images. (c) Lipids and MinD are colored in orange and cyan, respectively, by fluorescent dyes.
Scale bars: $5\,\mu$m.
(a) Reproduced from Ref.~\citenum{tame21} with permission from the Royal Society of Chemistry.
(b)--(c) Reproduced from Ref.~\citenum{lits18}.  Licensed under CC BY-NC.
}
\label{fig:vesos}
\end{figure}

\section{Reaction-Diffusion Dynamics on Membranes}\label{sec:reac}

In living cells, membrane remodeling often proceeds by multiple steps involving several proteins.
For example,
in clathrin-mediated endocytosis (CME),\cite{mcma11,mett18,kaks18}
first F-BAR-containing proteins and epsins bind on the membrane;
 subsequently, clathrins bind and assemble into form a spherical clathrin-coated bud.
N-BAR-containing proteins and dynamins bind to the bud neck, causing membrane fission.
Phosphoinositide synthesis and actin polymerization are also involved in CME.
Interactions among several proteins and other molecules often induce chemical waves and Turing-like patterns on membranes,
which can be modeled using reaction-diffusion equations.
The position of cell division in Escherichia coli is determined by the pole-to-pole oscillation of Min proteins.\cite{lutk12,meri21,hu99} 
The cell polarity of eukaryotic embryos and yeasts are determined by PAR\cite{gold07,lutk12,hoeg13,meri21} and Cdc42\cite{chio17,gory08} protein patterns, respectively.

In crawling cells, 
reaction-diffusion waves occur on the membrane to regulate 
cell locomotion.\cite{mogi20,beta23,alla13,arte14,inag17,pele11,tani13,imot21,tara22}
Many proteins and phosphoinositides are involved in these waves.
The membrane-bound WASP complex activates the Arp2/3 complex,
 which nucleates a new F-actin.
Other membrane-binding proteins, such as formins (e.g., mDia), promote F-actin growth.
Moreover, WASP and mDia are activated by phosphoinositide and GTPases such as Rac and Cdc42.
However, Arai et al. revealed that traveling waves of phosphoinositide are preserved even when actin polymerization is inhibited.\cite{arai10}
Hence, the actin is not necessary for wave generation.
They reproduced the observed wave patterns by a reaction-diffusion model for three components:
PIP$_3$ (phosphatidylinositol (3,4,5)-trisphosphate), PIP$_2$ (phosphatidylinositol (3,4,5)-bisphosphate),
and PTEN (phosphatase and ensin homolog).\cite{arai10}
PIP$_2$ and PIP$_3$ regulate the PTEN binding positively and negatively, respectively,
whereas PTEN promotes the dephosphorylation of PIP$_3$ to PIP$_2$.
This model has been extended to explain several experimental observations.
Fukushima et al.\cite{fuku19} added
the reaction-diffusion equations of Ras-GTP and RAS-GDP complexes to reproduce excitable waves in Ref.~\citenum{nish14}.
Taniguchi et al.\cite{tani13} added PI3K (phosphoinositide 3-kinase) to reproduce the dynamics of the cell shape.
A more general coupling of fast-excitable and slow-oscillatory systems was used in Ref.~\citenum{hsia13}.
These reaction waves can be accompanied by the membrane shape oscillation.
Wu et al. reported the oscillation of membrane height and F-BAR concentration in tumor cells.\cite{wu18}

Many components and reactions are involved in \textit{in vivo} wave generation;
hence, it is difficult to extract the effects of the membrane deformation.
To examine the coupling of membrane deformation and reaction-diffusion dynamics,
 we employed the reaction-diffusion equations of two components, $u$ and $v$.
One represents the concentration of a curvature-inducing protein on the membrane,  
and the other is the concentration of a regulatory protein or molecule.
Curvature-inducing proteins induce high bending rigidity and spontaneous curvature $C_0$,
while regulatory proteins do not directly modify the membrane properties.\cite{tame20,tame21,tame22,nogu23a}
The reaction-diffusion equations are as follows:
\begin{eqnarray}
\tau_{\mathrm {rd}}\frac{\partial u}{\partial t}&=& f(u,v) + D_{u} \nabla^2 u,  \\
\tau_{\mathrm {rd}}\frac{\partial v}{\partial t}&=& g(u,v) + D_{v} \nabla^2 v,
\end{eqnarray}
where $D_{u}$ and $D_{v}$ are diffusion constants and 
 $\tau_{\mathrm {rd}}$ is the reaction time unit.
As well-known reaction models,
the Brusselator\cite{prig68} and FitzHugh--Nagumo models\cite{fitz61,nagu62}
are employed with modifications for the coupling with the membrane deformation 
in Refs.~\citenum{tame20,tame21} and Refs.~\citenum{tame22,nogu23a}, respectively.
Membrane motion is solved by the Langevin equation using
a dynamically triangulated membrane model.\cite{nogu09,gomp04c,gomp97f,nogu05}
The membrane area $A$ and volume $V$ of a vesicle are fixed [i.e, the reduced volume $V^*= V/(4\pi {R_{\mathrm{ves}}}^{3}/3)$ is constant, where $R_{\mathrm{ves}}=(A/4\pi)^{1/2}$].

In Ref.~\citenum{tame20},
the binding rate of $u$ is linearly dependent on the bending energy change as follows:
\begin{eqnarray}
f(u,v) &=& A- G\frac{\partial f_{\mathrm {cv}}}{\partial u} - (B+1)u + u^2v,  \label{eq:f1} \\ \label{eq:f2}
g(u,v) &=& Bu - u^2v,
\end{eqnarray}
where $f_{\mathrm {cv}}$ is the local bending energy per unit area.
This is the standard Brusselator model at $G=0$ (no coupling).
As the membrane curvature approaches the preferred curvature,
the curvature-inducing protein ($u$) binds more frequently.
Figure~\ref{fig:turing} shows the phase diagram and typical vesicle shapes obtained in Ref.~\citenum{tame20}.
For  a non-deformable spherical vesicle,
the phase boundaries of the Turing pattern and temporal oscillation mode are determined by linear stability analysis 
[the Turing patterns appear between two solid lines in Fig.~\ref{fig:turing}(e)].
However, these conditions are modified by vesicle deformation.
The bound membranes (high $u$) are deformed in the direction of the preferred curvature,
whereas unbound membranes (low $u$) are often deformed in the opposite direction to maintain the volume [see Figs.~\ref{fig:turing}(a)--(d)].
Thus, the Turing patterns are stabilized, and the Turing-pattern region is enlarged in the phase diagram.
Budding and multi-spindle shapes are generated by the high-$u$ domains.
The hysteresis of vesicle shapes exists; initial oblate vesicles result in a larger number of spindles than prolate vesicles [see Figs.~\ref{fig:turing}(b) and (c)].
Spindles also increases with decreasing diffusion constants while keeping the ratio $D_{u}/D_{v}$, 
since the wavelength of the Turing patterns decreases.
For budded vesicles,
a Turing domain boundary separating two phases with high and low values of $u$ is formed at the connective neck, 
because protein diffusion is reduced at the narrow neck [see Fig.~\ref{fig:turing}(d)].
Moreover, a budding transition is observed to change a temporal oscillation of the protein concentration into a Turing pattern.

When wave propagation is sufficiently slow,
the membrane shape can largely oscillate in conjunction with the reaction-diffusion waves of curvature-inducing proteins,\cite{tame21,tame22}
and the re-entrant transition of wave disappearance can occur with increasing wave speed.\cite{nogu23a}
As an example, the spontaneous oscillation of a dumbbell-shaped vesicle is shown in Fig.~\ref{fig:vesos}(a).\cite{tame21}
Similar shape oscillations have been  experimentally observed for liposomes with a reconstituted Min system [see Fig.~\ref{fig:vesos}(b)--(c)].\cite{lits18,chri21}
Waves are produced by MinD, MinE, and ATP in \textit{in vitro} experiments.\cite{meri21,loos08,ramm18}
Traveling and standing waves are observed depending on geometry\cite{wu16,lits18,kohy19} and the ratios of MinD and MinE\cite{taka22} or ATP and dATP\cite{taka22a} experimentally.
These dynamics are reproduced by reaction-diffusion equations.\cite{case03,hala12,bonn13,wu16,kohy19,taka22} 
MinD-ATP complexes bind onto a membrane in an autocatalytic manner, and MinE is recruited to form the MinDE complex.
After the ATP hydrolysis, the MinD-ADP complex detaches from the membrane.
However, the influence of membrane deformation has not been included in these models, and
a further extension is required to take into account.

The shape oscillation of membranes has also been observed in liposomes involving microtubules and kinesin motors.\cite{kebe14} 
The microtubules form a nematic liquid crystal on the inner surface of the liposomes
and the kinesins slide the neighboring microtubules.
For spherical liposomes, the positions of $+1/2$ defects oscillate on the surface and 
are explained by the theoretical model, in which the defects move as self-propelled particles.
For lower reduced volumes, the microtubule bundles yield narrow membrane tube protrusions from the liposomes. The creation and annihilation of the protrusions cyclically occur.

\begin{figure}
\includegraphics[width=8.6cm]{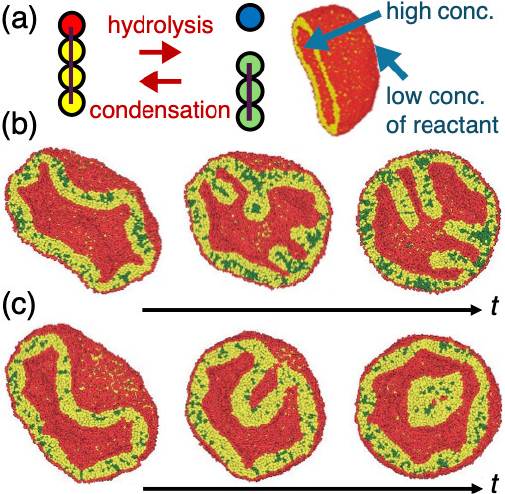}
\caption{
Shape transformation of vesicles induced by chemical reactions.
(a) Schematic of the hydrolysis and condensation reactions and an initial setup of DPD simulations.
An amphiphilic molecule is divided into hydrophilic and hydrophobic reactant molecules and vice versa.
In the initial state, a vesicle has a discocyte shape, and
the concentration of hydrophilic reactant molecules is higher inside than that outside.
(b) Sequential snapshots of bilayer sheet protrusion at high viscosity of surrounding fluids.
(c) Sequential snapshots of budding leading to vesicle formation at low fluid viscosity.
The hydrophilic and  hydrophobic segments of the amphiphilic molecules are displayed in red and yellow, respectively. 
The hydrophobic reactant molecules are in green.
The front halves of the vesicles, the hydrophilic reactant, and solvent molecules are not displayed for clarity.
Reproduced from Ref.~\citenum{naka18} with permission from the Royal Society of Chemistry.
}
\label{fig:sheet}
\end{figure}

\section{Membrane Dynamics Induced by the Synthesis and Decomposition of Amphiphilic Molecules}\label{sec:mol}

Until here,
we have reviewed the studies wherein the membrane maintains a bilayer structure
and is theoretically treated as an infinitely thin curved surface.
Reactions of lipids and other amphiphilic molecules can change the stability of their bilayer structure.
As the ratio of the hydrophilic portion increases,
the molecular aggregates change their shape from spherical,  cylindrical-like micelles, and bilayer membranes, to
inverted micelles.\cite{safr94,isra11}
The shape evolution from micelles to vesicles induced by chemical
reactions has been experimentally observed.\cite{suzu09,toyo06,taka13,blan11}
In experiments reported in Ref.~\citenum{toyo06}, 
amphiphilic molecules are divided into hydrophobic and hydrophilic molecules by hydrolysis,
and the resultant hydrophobic molecules 
decrease the spontaneous curvature of amphiphilic molecular 
assemblies, causing shape changes.
Initially, these amphiphilic molecules aggregate into a spherical micelle
structure. As the chemical reaction progresses, this subsequently
transforms into a tubular micelle, spherical vesicle, tubular vesicle, 
stomatocyte, and nested vesicle, eventually, forming oil droplets.
Kojima et al.\cite{koji23} reported the photo-induced transition between vesicles and droplets 
using azobenzene-containing amphiphiles, which reversibly isomerize in response to light.
Vesicle division by membrane growth has been observed using reactions 
to produce amphiphilic molecules.\cite{wald94,wick95,kuri11,cast19,miel20}
These division processes can be used for self-reproduction in protocells.\cite{goze22,imai22,howl23}

In molecular simulations, reactions can be treated by force fields and MC methods.
For atomistic molecular dynamics simulations,
a  force field called ReaxFF (reactive force field) has been developed to simulate bond breaking/formation based on quantum mechanics calculations.\cite{duin01,senf16,nayi23}
For coarse-grained molecular simulations,
MC methods are often used in combination with dissipation particle dynamics (DPD),
in which soft-repulsive potentials are employed.
Hence, polymerization\cite{lisa09,huan16,yan19,wang21} and condensation\cite{naka15,naka18} reactions have been simulated,
and vesicle formation has been observed.\cite{huan16,yan19,wang21,naka15}

These morphological changes in molecular assemblies can depend on the fluid properties in surrounding media.
Figure~\ref{fig:sheet} shows the morphological changes of vesicles by the formation and dissociation of amphiphilic molecules.\cite{naka18}
Since the initial concentration of hydrophilic reactant molecules is set to be higher inside than that outside,
the condensation reaction occurs more frequently in the inner leaflet, whereas the hydrolysis occurs in the outer leaflet.
As the reactions progress, the membrane area of the inner and outer leaflets increases and decreases, respectively.
When the fluid viscosity is relatively high compared to the membrane viscosity,
the bilayer sheets protrude inside the vesicle [see Fig.~\ref{fig:sheet}(b)].
Conversely, at low fluid viscosity, the vesicle exhibits a  budding into stomatocyte, leading to the formation of a small vesicle inside [see Fig.~\ref{fig:sheet}(c)].
High fluid viscosity slows vertical membrane motion more than lateral motion.

\section{Summary and Outlook}\label{sec:sum}

We have reviewed the recent studies on nonequilibrium membrane dynamics.
Membranes receive non-thermal forces from flows in channel proteins, binding and structural changes of proteins, and
interactions with protein filaments.
These forces modify the membrane fluctuations;
the non-thermal portion can be extracted as the violation of the fluctuation--dissipation relation.
Phase separation without the detailed balance between the phases can induce spatiotemporal patterns such as spiral waves.
Various membrane morphologies, such as Golgi-like stacked discocytes, can be generated by irreversible processes such as the fusion of small vesicles.
When the reaction-diffusion on a membrane is coupled with membrane deformation,
 chemical patterns and membrane shapes influence each other and cooperatively generate spatiotemporal patterns.
Chemical reactions that change the structure of amphiphilic molecules, such as the dissociation of amphiphilic molecules to hydrophilic and hydrophobic molecules, can alter the assembly structures among micelles, bilayers, and droplets. 

As explained, recent studies have progressed
our current knowledge on membrane dynamics.
However, many areas require further exploration.
New proteins and their functions are continually being discovered.
Since many proteins are usually involved in chemical waves in \textit{in vivo} experiments,
extracting the central players is often difficult.
Even if wave patterns are reproduced by a simulation of reaction-diffusion equations, other choices of equations may also provide a similar result. Careful examinations of parameter dependences in simulations and comparisons with experimental results
are important.
Recent development of machine learning and Bayesian optimization\cite{shah16} is very helpful for the model construction.
A Turing pattern and thermal microphase separation are not distinguishable only from a single spatial pattern.
Proteins can play opposite roles depending on conditions.
For example, actin filaments can bend the membrane by pushing it vertically but can also flatten the membrane by pulling it laterally.\cite{carl18}
We showed the positive feedback of membrane deformation to the reactions wherein the curvature-inducing proteins bind more frequently at the preferred curvature. This is a reasonable assumption when the curvature-inducing protein is a main player in reaction-diffusion equations.
However, the opposite (negative) feedback is possible when the binding rate of curvature-inducing proteins is predominantly determined by other proteins. 
Recent studies by Nishide and Ishihara\cite{nish22,nish24} have demonstrated that spatiotemporal patterns can be altered by the surface geometry even without surface deformation.
They have shown that pattern propagation occurs in axisymmetric surfaces in the absence of reflection symmetry, even on the condition that Turing patterns are stable on flat surfaces. The coupling with surface deformation can change dynamic modes further. 
The general understanding of coupling between reaction-diffusion dynamics and membrane deformation requires further investigations.

\begin{acknowledgments}
This work was supported by JSPS KAKENHI Grant Number JP24K06973.
\end{acknowledgments}

\end{document}